\documentstyle[12pt,fleqn,epsf]{article}
\def\npb#1#2#3{    {\it Nucl. Phys. }{\bf B #1} (19#2) #3}
\def\plb#1#2#3{    {\it Phys. Lett. }{\bf B #1} (19#2) #3}
\def\prd#1#2#3{    {\it Phys. Rev. }{\bf D #1} (19#2) #3}

\def\prl#1#2#3{    {\it Phys. Rev. Lett. }{\bf #1} (19#2) #3}

\def\mpla#1#2#3{   {\it Mod. Phys. Lett. }{\bf A #1} (19#2) #3}

\def\eq#1{{eq.~(\ref{#1})}}

\let\vev\VEV

\def\etal{{\it et al.}}

\newcommand{\bea}{\begin{eqnarray}}
\newcommand{\beq}{\begin{equation}}
\newcommand{\eea}{\end{eqnarray}}
\newcommand{\eeq}{\end{equation}}

\newcommand{\spav}[1]{\parbox{1mm}{\vspace*{#1}}}

\newcommand{\W}{{\scriptscriptstyle W}}
\newcommand{\HH}{{\scriptscriptstyle H}}
\newcommand{\KK}{{\scriptscriptstyle K}}

\newcommand{\alfasw}{\alpha_s({m_{\W}})}
\newcommand{\ep}{$\epsilon^\prime /\epsilon$~}
\newcommand{\epp}{$\epsilon^\prime /\epsilon$}

\begin{document}

\begin{titlepage}
\begin{flushright}
CERN-TH.7381/94\\
PREP N. 1016/94\\
\end{flushright}
\spav{0.0cm}
\begin{center}
{\Large\bf
Supersymmetric corrections to $\epsilon^\prime /\epsilon$ }\\
{\Large\bf
at the leading order in QCD and QED}\\
\spav{1.5cm}\\
{\large E. Gabrielli}
\spav{0.7cm}\\
{\em Dip. di Fisica, Universit\`{a} di Roma I ``La Sapienza''}\\
{\em P.le A. Moro 2, I-00185 Rome, Italy}\\
INFN -- Sezione di Roma I\\
\spav{1.0cm}\\
 {\large G.F. Giudice}\footnote{On leave of absence from INFN,
Sezione di Padova, Italy.}
\spav{0.7cm}\\
{\em CERN, Theory Division}\\
{\em CH-1211 Geneva 23, Switzerland}\\
\spav{1.5cm}\\
{\sc Abstract}
\end{center}
%%%%%%%%%%%%%%%%%%%%%%%%%%%%%%%%%%%%%%%%%%%%%%%%%%%%%%%%%%%%%%%%%%%%%%%%%%%%
%                            ABSTRACT
%%%%%%%%%%%%%%%%%%%%%%%%%%%%%%%%%%%%%%%%%%%%%%%%%%%%%%%%%%%%%%%%%%%%%%%%%%%%
We study the corrections to $\epsilon^\prime /\epsilon$ in the minimal
supersymmetric model at the leading order in QCD and QED. Supersymmetry
can increase the standard model prediction for $\epsilon^\prime
/\epsilon$ by at most 40\% for $m_t=174$ GeV, an enhancement which is
indistinguishable from the present theoretical uncertainties.
The most conspicuous effect of supersymmetry is a strong depletion
of $\epsilon^\prime /\epsilon$. For certain choices of supersymmetric
parameters, vanishing and even small negative values of $\epsilon^\prime
/\epsilon$ can be obtained for the top quark in the CDF range.
 \vfill
\spav{.5cm}\\
CERN-TH.7381/94\\
PREP. N. 1016/94\\
 July 1994
\end{titlepage}

\newpage
\setcounter{footnote}{0}
\setcounter{page}{1}

\bigskip
%%%%%%%%%%%%%%%%%%%%%%%%%%%%%%%%%%%%%%%%%%%%%%%%%%%%%%%%%%%%%%%%%%%%%%%%%%%%
%                       INTRODUCTION
%%%%%%%%%%%%%%%%%%%%%%%%%%%%%%%%%%%%%%%%%%%%%%%%%%%%%%%%%%%%%%%%%%%%%%%%%%%%
\section{Introduction}
CP violation in the kaon system has always been a fertile field in which
to test theories beyond the Standard Model (SM). At present the situation
for \ep is particularly interesting. The experimental results from NA31
at Cern \cite{na31}
\beq
\mbox{Re}~\epsilon^\prime /\epsilon =23\pm 6.5 \times 10^{-4}
\label{na31q}
\eeq
and from E731 at Fermilab \cite{e731}
\beq
\mbox{Re}~\epsilon^\prime /\epsilon =7.4\pm 6.0 \times 10^{-4}
\label{e731q}
\eeq
are incompatible at the 1-$\sigma$ level. On the theoretical side,
there has been great progress in reducing the uncertainties in the
prediction
for \epp. The full next-to-leading QCD perturbative corrections have
been computed by two groups \cite{nlob,nlom} and developments in lattice
calculations should soon provide reliable estimates of the hadronic
matrix elements.

Given the inconclusive experimental situation and the notable refinement
in the SM calculation for \epp, the time is right for detailed investigations
of \ep in theories beyond the SM. If the top quark is indeed as heavy
as implied by the CDF measurements \cite{cdf}, the next-to-leading
SM prediction for \ep sits comfortably in the E731 range,
disfavoring the larger NA31 result. While we await
improved experimental statistics, which will resolve the conflict between
\eq{na31q} and \eq{e731q}, it is now interesting to
see if extensions of the SM can substantially modify the prediction for \ep
and possibly account for the large value suggested by the NA31 measurement.
Furthermore, since the SM prediction for \ep suffers a strong accidental
cancellation among the different contributions for the top
quark in the range of interest, it is conceivable that even modest
new-physics effects will easily stand out and sizeably modify the final
result.

A notable example of study of \ep in theories beyond the SM is contained
in ref.\cite{burh}. There, in the context of the two-Higgs doublet model,
a systematic analysis of \ep including leading QCD logarithms has been
carried out. It was found that the charged Higgs has the effect of reducing
\ep with respect to the SM prediction. This reduction occurs partly
because of a positive charged-Higgs contribution to $\epsilon$, which
suppresses the Cabibbo-Kobayashi-Maskawa (CKM) CP violating parameter
$\sin \delta$, and partly because of a new contribution to electroweak
penguin and box diagrams, which enhances the degree of cancellation
with the strong penguin already present in the SM for a heavy top
quark.

The aim of this paper is to generalize the analysis of ref.\cite{burh}
to the case of the minimal supersymmetric model. Although several
analyses of \ep in the context of supersymmetry are already present in the
literature \cite{epssusy}, a complete study including leading QCD
logarithms has not been performed. In order to attempt such a study,
it is necessary to make precise assumptions regarding the structure
of CP violation and flavor-changing neutral currents (FCNC) in the
supersymmetric model. Here we choose to work in a minimal version of the
model, described in more detail in sect.~2, where both CP violation
and FCNC are completely determined by the usual CKM matrix elements.
We have several reasons to do this:

{\it i)} The corrections to \ep considered here are fairly generic to all
supersymmetric models. Non-minimal models may contain new CP-violating
phases and/or tree-level FCNC, thereby generating extra
contributions to \epp. These contributions are however strongly
model-dependent.

{\it ii)} The minimal version of the supersymmetric model considered here
is very predictive, since CP violation and FCNC are completely determined
in terms of only the known CKM angles and phases.

{\it iii)} The effective Hamiltonian below the weak scale can be written
in terms of the same set of operators used in the case of the SM. This
greatly simplifies the analysis, since the anomalous dimension matrix
is then completely known at the leading \cite{lo} and next-to-leading
\cite{nlo} order. If, for instance, flavor-changing gluino-mediated
interactions were included, new operators with different chiral
structure would appear and a calculation of new elements of
the anomalous-dimension matrix would be required.

Our paper is organized as follows. In sect.~2 we describe the version
of the supersymmetric model under investigation and we establish our
notation. In sects.~3 and 4 we give the formulae for the supersymmetric
corrections to $\epsilon$ and \epp. The results of our numerical
investigation are presented in sect.~5. Finally, in sect.~6 we
summarize our results.

%%%%%%%%%%%%%%%%%%%%%%%%%%%%%%%%%%%%%%%%%%%%%%%%%%%%%%%%%%%%%%%%%%%%%%%%%%%%
%     SUPERSYMMETRIC MODEL WITH MINIMAL CP AND FLAVOR VIOLATION
%%%%%%%%%%%%%%%%%%%%%%%%%%%%%%%%%%%%%%%%%%%%%%%%%%%%%%%%%%%%%%%%%%%%%%%%%%%%

\section{Supersymmetric model with minimal CP and flavor violation}

In general, supersymmetric models have potential sources of new CP
violation because of the presence of unremovable phases in the
supersymmetry-breaking terms. However, these phases are strongly
constrained by measurements on the electric dipole moment of
the neutron and must be small, typically less than $10^{-2}-10^{-3}$
\cite{epssusy}. Although  there is no compelling theoretical argument
which suggests that they are exactly zero, we will make the simplifying
(and often used)
assumption that the only CP violation resides in the CKM matrix.

It is also well known that FCNC in
supersymmetric models can arise at the tree level, since the quark-
and squark-mass matrices are diagonalized by
different field transformations \cite{fcnc}.
Once the heavy supersymmetric particles are integrated out, one is left with
FCNC involving ordinary quarks, which are induced by one-loop diagrams with
squark and gluino exchange. Their effects are potentially large, since
they are O($\alpha_s^2$), as opposed to the standard model FCNC contributions
O($\alpha_W^2$). The flavor-changing quark-squark-gluino couplings
depend however on the
details of the supersymmetry-breaking sector, which is still the
least-understood part of the theory. It is customary to make the
simplifying assumption that all supersymmetry-breaking terms are
flavor independent at some grand-unification scale, close to the Planck mass.
This is the case if, for instance, the K\"ahler metric of the underling
supergravity theory is flat.
In this case flavor-changing quark-squark-gluino couplings are induced
by renormalization effects, but their influence in kaon physics
is negligible because
of the strong limits on gluino and squark masses
  from hadron colliders \cite{cdfsusy}.

There is growing criticism of this point of view
\cite{kap}, since supergravity
theories derived from superstrings do not seem to have flat K\"ahler
metrics and do not seem to have flavor-independent
supersymmetry-breaking terms. However, if the supersymmetry-breaking terms
have a completely general structure in flavor space, they lead to corrections
to $K^0-\bar{K}^0$ mixing larger than that
allowed by experimental constraints.
In this case there is need for a suppression of the flavor-changing
quark-squark-gluino
coupling either by postulating some form of flavor symmetry at the unification
scale or by invoking some dynamical mechanism to make squarks
more degenerate in mass ({\it e.g.} the
running from unification to weak scale
in a gaugino-dominated supersymmetry-breaking scenario).

In our analysis we will not be concerned about the fundamental
mechanism which suppresses FCNC, but we will simply postulate,
as we have done for CP violation,
that, at the weak scale, FCNC are
absent at tree level. As discussed in
the introduction, this hypothesis of minimal CP and flavor violation
enhances the predictivity of the model and largely simplifies the
analysis of QCD effects, since the operator basis of the effective
Hamiltonian is the same as in the SM.
Let us now briefly present the theoretical framework
in which we work, the supersymmetric standard model with minimal
CP and flavor
violation, and establish our notation.

By making a transformation on superfields, we choose a basis in which
the up-quark mass matrix $m_u$ is real and diagonal and the down-quark
mass matrix is $Vm_d$, where $m_d$ is real and diagonal and $V$ is the
CKM matrix. In this basis, the $6\times 6$ up- and down-squark mass matrices
are:
\beq
\tilde{M}^2_u=\left(
\matrix{
\tilde{m}_{u_L}^2+m_u^2 & A_um_u \cr
A_um_u &\tilde{m}_{u_R}^2+m_u^2 \cr}
\right) ,
\label{squarkmassu}
\eeq
\beq
\tilde{M}^2_d=\left(
\matrix{
\tilde{m}_{d_L}^2+Vm_d^2V^{\dag} & A_dVm_d \cr
A_dm_dV^{\dag}&\tilde{m}_{d_R}^2+m_d^2 \cr}
\right) .
\label{squarkmassd}
\eeq
Our minimal CP- and flavor-violation hypothesis states that,
at the weak scale,
the supersym- metry-breaking terms $\tilde{m}_{u_{L,R}},~\tilde{m}_{d_{L,R}}$
and $A_{u,d}$ are CP-conserving and flavor-independent, {\it i.e.}
real and proportional to the $3\times 3$
identity matrix. If this hypothesis
holds, there are no tree-level flavor-violating
quark-squark-gluino couplings at the weak scale. Because
of the strong constraints imposed
by the real part of $K^0-\bar{K}^0$ mixing, we know that this hypothesis
should be approximately satisfied, unless squarks or gluinos are very heavy.
\par
The mass matrices eqs. (\ref{squarkmassu})--(\ref{squarkmassd})
can be diagonalized according to
\beq
T_{u}\tilde{M}^2_uT^{\dag}_u=\mbox{diag}\left(
\frac{\tilde{m}_{u_{Li}}^2
+\tilde{m}_{u_{Ri}}^2}{2}+
m^2_{u_i}+
\frac{A_um_{u_i}}{\sin{2\theta^u_i}},
\frac{\tilde{m}_{u_{Li}}^2
+\tilde{m}_{u_{Ri}}^2}{2}+
m^2_{u_i}-\frac{A_um_{u_i}}{\sin{2\theta^u_i}}\right) ,
\eeq
\beq
T_{d}\tilde{M}^2_dT^{\dag}_d=\mbox{diag}\left(
\frac{\tilde{m}_{d_{Li}}^2
+\tilde{m}_{d_{Ri}}^2}{2}+
m^2_{d_i}+
\frac{A_dm_{d_i}}{\sin{2\theta^d_i}},
\frac{\tilde{m}_{d_{Li}}^2
+\tilde{m}_{d_{Ri}}^2}{2}+
m^2_{d_i}-\frac{A_dm_{d_i}}{\sin{2\theta^d_i}}\right) ,
\eeq
with $i=1,2,3$ and where
\beq
T_u=
\left(
\matrix{
C_u & S_u \cr
-S_u & C_u \cr}
\right),~~~
T_d=
\left(
\matrix{
C_dV^{\dag} & S_d \cr
-S_dV^{\dag} & C_d \cr}
\right) ,
\eeq
\bea
C_{u,d}&=& \mbox{diag}\left(\cos{\theta_i^{u,d}}\right) ,~~~~~
S_{u,d}= \mbox{diag}\left(\sin{\theta_i^{u,d}}\right) ,\\
\tan{2\theta_i^{u,d}}&=&\frac{2A_{u,d}m_i^{u,d}}{
\tilde{m}_{u_{Li},d_{Li}}^2-
\tilde{m}_{u_{Ri},d_{Ri}}^2} .
\eea

The chargino mass matrix is diagonalized by two orthogonal
$2\times 2$ matrices
$U$ and $V$, according to:
\beq
U\left(
\matrix{
M & m_{\W}\sqrt{2}\sin{\beta} \cr
m_{\W}\sqrt{2}\cos{\beta} & \mu \cr}
\right)V^{-1}=
\left(
\matrix{
\tilde{m}_{\chi_1} & 0 \cr
0 & \tilde{m}_{\chi_2} \cr}
\right) ,
\nonumber
\eeq
where $\tan{\beta}$ is the ratio of the two Higgs vacuum expectation
values, $M$ is the weak gaugino mass, and $\mu$ is the Higgs superfield
mixing parameter, taken here to be real.

In order to compute the relevant one-loop diagrams, we need the
interaction of charginos $(\tilde{\chi}_j,~j=1,2)$ with down quarks
($d)$ and up squarks $(\tilde{u}_k,~k=1,2)$ in terms of mass eigenstates:
\beq
{\cal L}_{\chi}=g\bar{d}V^\dagger \left(
Z^{jk}\frac{1-\gamma_5}{2}+
Y^{jk}\frac{1+\gamma_5}{2}\right)\tilde{\chi}_j^{(-)}
\tilde{u}_k+\mbox{h.c.},
\label{Lagcharg}
\eeq
\bea
Z^{jk}&=&\frac{m_d}{\sqrt{2}m_{\W}\cos{\beta}}U_{j2}
\left(
\begin{array}{c}
\cos{\theta^u} \\
-\sin{\theta^u} \\
\end{array}
\right)_k\\
Y^{jk}&=&\frac{m_u}{\sqrt{2}m_{\W}\sin{\beta}}V_{j2}
\left(
\begin{array}{c}
\sin{\theta^u}\\
\cos{\theta^u}\\
\end{array}
\right)_k
-V_{j1}
\left(
\begin{array}{c}
\cos{\theta^u}\\
-\sin{\theta^u}\\
\end{array}
\right)_k ,
\eea
where flavor indices are understood. The CP- and flavor-violation
properties of the interaction Lagrangian in
\eq{Lagcharg} are determined by the familiar CKM matrix.
This property is satisfied also by
the charged-Higgs interactions with quarks:
\beq
{\cal L}_{\HH}=\frac{g}{\sqrt{2}m_{\W}}H^+
\bar{u}\left(\frac{1}{\tan \beta}m_uV
\frac{1-\gamma_5}{2}+\tan \beta Vm_d
\frac{1+\gamma_5}{2}\right) d
+\mbox{h.c.}
\eeq

We conclude this section by summarizing the free parameters necessary to
specify the supersymmetric model, in addition to those of the SM.
Whenever convenient, we choose to redefine the original parameters and work
with physical particle masses as input. For the first two
generations of up-squarks $m_u$ is negligible with respect to
$\tilde{m}_{u_{L,R}}$ and we will take these squarks to be degenerate in mass,
i.e. $\tilde{m}_{u_L}\simeq\tilde{m}_{u_R}\equiv\tilde{m}$.
This is merely a simplifying assumption which will not affect our conclusions.
The mixing between the two stops cannot be neglected. We choose as input
parameters the lightest stop mass, $m_{\tilde{t}}$, and the stop mixing
angle $\theta$ (which can be chosen in the range between 0 and $\pi$,
without loss of generality). The mass of the heavier stop ($m_{\tilde{T}}$)
is then given by:
\beq
m_{\tilde{T}}^2=2\tilde{m}+2m_t^2-m_{\tilde{t}}^2.
\nonumber
\eeq
The chargino mass matrix is defined by three parameters: we choose them
to be the lightest chargino mass ($m_{\chi}$), the ratio of Higgs vacuum
expectation values ($\tan{\beta}$), and the weak gaugino mass ($M$).
The last free parameter is the charged-Higgs mass ($M_{H^{+}}$),
while the relevant $H^{\pm}$ couplings are completely determined once
the value of $\tan{\beta}$ is fixed. Therefore the model requires seven
input parameters: $\tilde{m},~m_{\tilde{t}},~\theta,~m_{\chi},~\tan{\beta},
{}~M,~M_{H^+}$. Notice that some of these seven free parameters may become
related with one other if additional assumptions are made (specific
relations at the GUT scale, constraints from electroweak symmetry breaking,
etc.). We prefer to treat them as independent variables in order to
describe a larger class of models.
%%%%%%%%%%%%%%%%%%%%%%%%%%%%%%%%%%%%%%%%%%%%%%%%%%%%%%%%%%%%%%%%%%%%%%%%%%%%
%                     THE PARAMETER EPS
%%%%%%%%%%%%%%%%%%%%%%%%%%%%%%%%%%%%%%%%%%%%%%%%%%%%%%%%%%%%%%%%%%%%%%%%%%%%

\section{The parameter $\epsilon$ }

The parameter $\epsilon$ gives a measure of the indirect CP-violation
in kaon decays. It is defined as
\beq
\epsilon=\frac{e^{i\pi/4}}{\sqrt{2}\Delta m_K}\mbox{Im}M_{12},
\label{eps}
\eeq
where $M_{12}$ is the imaginary part of
the off-diagonal element of the neutral kaon
system $(K^0,\bar{K}^0)$ mass matrix, and
$\Delta m_K$ is the $K_L-K_S$ mass difference.

By evaluating the relevant $\Delta S=2$ box diagrams in figure 1a-b, we obtain
\bea
\epsilon&=&e^{i\pi/4}\frac{G_F^2}{12\sqrt{2}\pi^2}f^2_{\KK}
\frac{m_{\KK}}{\Delta m_{\KK}}B_K m^2_{\W}\sum_i \left[
(\mbox{Im} \lambda_c^2)\eta_1^{(i)}S^{(i)}(c,c)+
\right.
\nonumber \\
&&
\left.
(\mbox{Im} \lambda_t^2)\eta_2^{(i)}S^{(i)}(t,t)+
2(\mbox{Im} \lambda_c\lambda_t)\eta_3^{(i)}S^{(i)}(c,t)
\right]
\label{epsf}
\eea
where $\lambda_i=V_{id}^* V_{is}$, $f_\KK =161$ MeV is the
kaon decay constant, and $B_K$ is the non-perturbative
parameter which defines the normalization of the relevant hadronic
matrix element in units of the vacuum-insertion value.
The coefficients $\eta_i$ include the QCD corrections to the box diagrams
and are given in table 1. The values of $\eta_i$ for the
box diagrams involving squarks and charginos are all equal to the
SM value of $\eta_2$ because we are assuming that the supersymmetric
particles are integrated out at the scale of the $W$ boson.

The functions $S^{(i)}(c,t)$ result from calculation of the
box diagrams with exchange of quarks and two W bosons ($WW$),
two charged Higgs bosons ($HH$), one W and one charged Higgs ($HW$),
and with exchange of squarks and charginos ($\chi$):
\bea
S^{\W\W}(c,t)&=&S(x_{c\W},x_{t\W})\nonumber \\
S^{\HH\HH}(c,t)&=&\frac{x_{\HH\W}}{4\tan^4{\beta}}
L_2(x_{c\W},x_{t\W},1)\nonumber \\
S^{\HH\W}(c,t)&=&\frac{2}{\tan^2{\beta}}\left[\frac{1}{4}L_2(x_{c\W},
x_{t\W},x_{\HH\W})-
L_1(x_{c\W},x_{t\W},x_{\HH\W})\right]\nonumber \\
S^{\chi}(c,t)&=&f(u,u)-f(u,c)-f(u,t)+f(c,t)\nonumber \\
f(c,t)&=&
\sum_{\stackrel{i,j=1,2}{\scriptstyle{h,k=1,2}}}
\frac{x_{\W\chi_j}}{4}
Y_{i\tilde{c}_h}Y_{i\tilde{t}_k}
Y_{j\tilde{c}_h}Y_{j\tilde{t}_k}
L_3(x_{\tilde{t}_{k}\chi_j},x_{\tilde{c}_{h}\chi_j},x_{\chi_i\chi_j}) .
\label{fboxs2}
\eea
We have denoted
$x_{ab}\equiv m_a^2/m_b^2$ for generic indices $a,b$, and have collected
the functions $S$ and $L_{1,2,3}$ in the appendix.
The expressions for the charged-Higgs contributions
to short-distance effects given in this and the next section
agree with ref. \cite{burh}, and the expressions
for the chargino contributions agree with the existent literature
whenever results are available (see, in particular, ref.\cite{bbmr}).

For a given set of supersymmetric parameters, one can compare the
theoretical prediction for $\epsilon$ in \eq{epsf} with the experimental
result and then determine $\delta$, the CP-violating phase in the
``standard" representation \cite{rpd} of the CKM matrix. In general
one obtains two solutions for $\delta$ but, in most of the parameter
space, one of them is ruled out by the experimental data on $B^0-\bar{B}^0$
mixing and on the $b\to s \gamma$ branching ratio.

The $B^0-\bar{B}^0$ mixing is determined by the box diagrams in figure
1c-d in a manner analogous to $\epsilon$:
\beq
x_d\equiv \frac{\Delta m_B}{\Gamma_B}=\tau_B\frac{G_F^2}{6\pi^2}
\eta_{QCD} m_B B_B f_B^2 m_{\W}^2 |V_{td}|^2\sum_i S^{(i)}(t,t).
\eeq
We take $m_B=5.28$ GeV, the perturbative QCD correction parameter
$\eta_{QCD}=0.84$, and the experimental determination
   from ARGUS \cite{arg} and CLEO \cite{cleo} for $x_d=0.70\pm0.13.$

The expression for the decay rate of $b\to s \gamma$ in supersymmetry
can be found in ref. \cite{bbmr} (or, in a notation very similar
to the one employed here, in ref. \cite{giu}) and will not be repeated
here. In our analysis we will impose the constraint on the inclusive
branching ratio $BR(b\to s\gamma )
<5.4 \times 10^{-4}$ obtained by CLEO \cite{cleobsg}. The numerical
results will be presented in sect. 5.
%%%%%%%%%%%%%%%%%%%%%%%%%%%%%%%%%%%%%%%%%%%%%%%%%%%%%%%%%%%%%%%%%%%%
%                   THE PARAMETER EPS'
%%%%%%%%%%%%%%%%%%%%%%%%%%%%%%%%%%%%%%%%%%%%%%%%%%%%%%%%%%%%%%%%%%%%
\section{The parameter $\epsilon^\prime$}
The parameter $\epsilon^\prime$ gives a measure of the direct CP-violation
in kaon decays.
It is defined as follows:
\beq
\epsilon^\prime =-\frac{\omega}{\sqrt{2}}\xi \left(1-\Omega\right)
              e^{i\tilde{\phi}},
\label{epsp}
\eeq
where
\beq
\xi=\frac{\mbox{Im}A_0}{\mbox{Re}A_0},~~~~~
\omega=\frac{\mbox{Re}A_2}{\mbox{Re}A_0},~~~~~
\Omega=\frac{1}{\omega}\frac{\mbox{Im}A_2}{\mbox{Im}A_0},~~~~~
\tilde{\phi}=\frac{\pi}{2}+\delta_2-\delta_0 \simeq \frac{\pi}{4}.
\eeq
$A_I$ and $\delta_I$ are the amplitudes and final state interaction phases
for the decay $K\to (\pi \pi )_I$, where $I$ denotes the isospin of the
final pion state. Experimentally one has
\beq
\mbox{Re}A_0=3.3\times 10^{-7} \mbox{GeV},~~~~~~~~\omega\simeq 1/22.
\label{rea0}
\eeq
The procedure to relate the amplitudes in \eq{epsp} to the QCD-improved
$\Delta S=1$ effective Hamiltonian is by now standard \cite{effec,bur}
We will follow the method illustrated in ref. \cite{bur}, briefly
review its results, and show where modifications due to the
supersymmetric contributions are necessary.

In the SM, after integrating out the heavy degrees of freedom at the
scale of the $W$ boson,
the tree-level $\Delta S=1$ Hamiltonian becomes:
\beq
H_{eff}^{\Delta S=1}=\frac{G_F}{\sqrt{2}}V_{ud}V^{*}_{us}\left[
\left(1-\tau\right)\left( Q^u_2-Q^c_2\right)+
\tau\left( Q^u_2-Q^t_2\right)\right] ,
\label{hs1}
\eeq
where
\beq
\tau=-\frac{V_{td}V^{*}_{ts}}{V_{ud}V^{*}_{us}},~~~~~
Q_2^{q}=\left(\bar{s}q\right)_{V-A}
\left(\bar{q}d\right)_{V-A},~~q=u,c,t.
\label{q2i}
\eeq
With QCD corrections taken into account,
the effective Hamiltonian at an energy scale $\mu$ below the
charm mass is obtained by replacing in \eq{hs1}
the operators $Q_2^q$
with the renormalized four-fermion operators $Q_i(\mu)$ as follows:
\beq
\left( Q_2^u-Q_2^c\right)\rightarrow\sum_{i=1}^{10}z_i(\mu)Q_i(\mu),~~~~~~~~
\left( Q_2^u-Q_2^t\right)\rightarrow\sum_{i=1}^{10}v_i(\mu)Q_i(\mu).
\label{base}
\eeq
In \eq{base},
$v_i(\mu)$ and $z_i(\mu)$ correspond to the relevant Wilson coefficients
and the  complete basis used for the operators $Q_i$ is
\bea
Q_1&=&(\bar{s}d)_{V-A}(\bar{u}u)_{V-A},
{}~~~~~~~~~~~~~~Q_2~=~(\bar{s}u)_{V-A}(\bar{u}d)_{V-A},
\nonumber\\
Q_3&=&(\bar{s}d)_{V-A}\sum_{q=u,d,s}(\bar{q}q)_{V-A},
{}~~~~~~~~~Q_4~=~\sum_{q=u,d,s}(\bar{s}q)_{V-A}(\bar{q}d)_{V-A},
\nonumber\\
Q_5&=&(\bar{s}d)_{V-A}\sum_{q=u,d,s}(\bar{q}q)_{V+A},
{}~~~~~~~~~Q_6~=~-8\sum_{q=u,d,s}(\bar{s}_Lq_R)(\bar{q}_Rd_L),
\nonumber\\
Q_7&=&\frac{3}{2}(\bar{s}d)_{V-A}\sum_{q=u,d,s}e_q(\bar{q}q)_{V+A},
{}~~~~~Q_8~=~-12\sum_{q=u,d,s}e_q(\bar{s}_Lq_R)(\bar{q}_Rd_L),
\nonumber\\
Q_9&=&\frac{3}{2}(\bar{s}d)_{V-A}\sum_{q=u,d,s}e_q(\bar{q}q)_{V-A},
{}~~~~Q_{10}~=~\frac{3}{2}\sum_{q=u,d,s}e_q(\bar{s}q)_{V-A}(\bar{q}d)_{V-A},
\label{qibase}
\eea
where $e_q$ is the electric charge of the quark $q$,
$(V\pm A)$ refer to $\gamma_{\mu}(1\pm\gamma_5)$, and
$q_{L,R}=1/2(1\pm\gamma_5)q$.
We neglect here magnetic-moment operators \cite{bfg}, since their
contribution is unimportant for our purposes.

In general we can therefore write the effective $\Delta S=1$
Hamiltonian at the scale $\mu$ as
\beq
H_{eff}^{\Delta S=1}=\frac{G_F}{\sqrt{2}}V_{ud}V^{*}_{us}
\sum_{i=1}^{10}
C_i(\mu )Q_i(\mu),
\label{hs1mu}
\eeq
where
\beq
C_i(\mu)= z_i(\mu)+\tau y_i(\mu);~~~~~y_i(\mu)\equiv v_i(\mu)-z_i(\mu).
\label{yi}
\eeq
The Wilson coefficients $C_i$ are chosen to match the short-distance
contribution at the appropriate scale where heavy particles are
integrated out, and then evolved using the renormalization group equation:
\beq
\left[\mu\frac{\partial}{\partial\mu}+\beta(g)\frac{\partial}{\partial g}-
\hat{\gamma}^T(g^2,\alpha)
\right]\vec{C}\left(\frac{m_{\W}^2}{\mu^2},g^2,\alpha\right)=0,
\label{rge}
\eeq
where $\vec{C}$ stands for a ten-dimensional vector, $\beta(g)$
is the QCD beta function, and $\alpha$ is the
electromagnetic coupling constant (the running of $\alpha$ is neglected).
In this paper we use the anomalous-dimension matrix
$\hat{\gamma}^T(g^2,\alpha) $ at the leading order in QCD and QED as given in
ref. \cite{bur}, although it is now known at the next-to-leading order
\cite{nlo}. This is perfectly adequate, in view of the large
intrinsic uncertainty due to the ignorance of the supersymmetric
parameters and to the inescapable model-dependence. Our aim here
is just to illustrate the trend of the supersymmetric corrections
to \epp.

The modifications caused by
supersymmetry appear only in the boundary conditions
of the Wilson coefficients, which are computed through the appropriate
one-loop Feynman diagrams in figure 2. In our analysis we will impose the
boundary conditions at the scale $\mu =m_{\W}$, although strictly
speaking they should apply to the energy scale at which the heavy
particles are integrated out. The threshold corrections,
originating from this mismatch of energy scales and certainly
present in any realistic theory with a non-degenerate mass-spectrum, are
numerically not very significant, since the running of $\alpha_s$ in
the region above $m_{\W}$ is not very steep. In the HV renormalization scheme,
the boundary conditions for
the Wilson coefficients in the minimal supersymmetric model are given by:
\bea
v_1(m_{\W})&=&\frac{\alfasw}{16\pi}\left(14-B_{\tilde{g}}\right)\nonumber \\
v_2(m_{\W})&=& 1\nonumber \\
v_3(m_{\W})&=& \frac{\alpha}{6\pi}\frac{1}{\sin^2{\theta_{\W}}}\left(
B^{(d)}+\frac{B^{(u)}}{2}+C\right)-\frac{\alfasw}{24\pi}E\nonumber \\
v_4(m_{\W})&=& \frac{\alfasw}{8\pi}E\nonumber \\
v_5(m_{\W})&=& -\frac{\alfasw}{24\pi}E\nonumber \\
v_6(m_{\W})&=& \frac{\alfasw}{8\pi}E\nonumber \\
v_7(m_{\W})&=&\frac{\alpha}{6\pi}\left(4C+D\right)\nonumber \\
v_8(m_{\W})&=& 0\nonumber \\
v_9(m_{\W})&=& \frac{\alpha}{6\pi}\left[4C+D+\frac{1}{\sin^2{\theta_{\W}}}
\left(-B^{(d)}+B^{(u)}-4C\right)\right]\nonumber \\
v_{10}(m_{\W})&=& 0
\label{vimw}
\eea
and
\beq
z_1(m_{\W})=v_1(m_{\W}),~~z_2(m_{\W})=v_2(m_{\W}),~~z_i(m_{\W})=0,~~i=3,10.
\label{zimw}
\eeq
The functions which include the contributions from photon-penguins ($D$),
Z-penguins ($C$), gluon-penguins ($E$), boxes with external down
quarks ($B^{(d)}$) and up quarks ($B^{(u)}$), and gluino-mediated boxes
($B^{({\tilde{g}})}$) are given by:
%
%  PHOTON PENGUIN
%
\beq
D=D_{SM}(x_{t\W})+\frac{1}{\tan^2\beta}D_H(x_{t\HH})+
\sum_{\stackrel{j=1,2}{\scriptstyle{k=1,2}}}
\left[
Y^2_{j\tilde{t}_k}x_{\W\tilde{t}_k}D_{\chi}(x_{\chi_j \tilde{t}_k})-
\left(\tilde{t}\rightarrow \tilde{c}\right)\right]
\label{fpeng}
\eeq
%
%  Z PENGUIN
%
\bea
C&=&C_{SM}(x_{t\W})+\frac{x_{t\W}}{\tan^2\beta}C_H(x_{t\HH})+
\sum_{\stackrel{i,j=1,2}{\scriptstyle{h,k=1,2}}}
\left[
Y_{j\tilde{t}_{k}}Y_{i\tilde{t}_{h}}
\left\{
\frac{1}{2}\delta_{ij}\Delta_{hk}
C_{\chi}^{(1)}(x_{\tilde{t}_h\chi_j},x_{\tilde{t}_k\chi_j})
\right.\right.
\nonumber \\
&+&
\left.\left.
\delta_{hk}\left[
U_{i1}U_{j1}C^{(2)}_{\chi}(x_{\chi_j\tilde{t}_k},x_{\chi_i\tilde{t}_k})+
V_{i1}V_{j1}\left(\frac{1}{16}\log{m^2_{\tilde{t}_k}}-
C^{(1)}_{\chi}(x_{\chi_j\tilde{t}_k},x_{\chi_i\tilde{t}_k})
\right)\right]\right\}
\right.
\nonumber \\
&-&
\left.
\left(\tilde{t}\rightarrow\tilde{c}\right)\right]
\label{zpeng}
\eea
%
%  GLUON PENGUIN
%
\beq
E=E_{SM}(x_{t\W})+\frac{1}{\tan^2\beta}E_{H}(x_{t\HH})+
\sum_{\stackrel{j=1,2}{\scriptstyle{k=1,2}}}
\left[Y^2_{j\tilde{t}_{k}}x_{\W\tilde{t}_k}
E_{\chi}(x_{\chi_j\tilde{t}_k})-\left(\tilde{t}\rightarrow\tilde{c}\right)
\right]
\label{gpeng}
\eeq
%
%  BOX (d) CHARGINI (DELTA S)=1
%
\bea
B^{(d)}&=&-2B_{SM}(x_{t\W})+
\frac{1}{8}\sum_{\stackrel{i,j=1,2}{\scriptstyle{h,k=1,2}}}
Y_{j\tilde{u}_{h}}Y_{i\tilde{u}_{h}}x_{\W\chi_j}\left[
Y_{j\tilde{t}_{k}}Y_{i\tilde{t}_{k}}
B_{\chi}^{(d)}(x_{\tilde{t}_k\chi_j},x_{\tilde{u}_h\chi_j},x_{\chi_i\chi_j})
\right.\nonumber \\
&-&
\left.
\left(\tilde{t}\rightarrow\tilde{c}\right)\right]
\label{dbox}
\eea
%
%  BOX (u) CHARGINI (DELTA S)=1
%
\bea
B^{(u)}&=&8B_{SM}(x_{t\W})+
\frac{1}{4}\sum_{\stackrel{i,j=1,2}{\scriptstyle{h,k=1,2}}}
U_{j1}U_{i1}x_{\W\chi_j}\sqrt{x_{\chi_i\chi_j}}\left[
Y_{j\tilde{t}_{k}}Y_{i\tilde{t}_{k}}
B_{\chi}^{(u)}(x_{\tilde{t}_k\chi_j},x_{\tilde{d}_h\chi_j},x_{\chi_i\chi_j})
\right. \nonumber \\
&-&
\left.
\left(\tilde{t}\rightarrow\tilde{c}\right)\right]
\label{ubox}
\eea
%
%  BOX gluini
%
\bea
B^{({\tilde{g}})}&=&\sum_{\stackrel{i=1,2}{\scriptstyle{h,k=1,2}}}
\Delta_{hk}x_{\W g}\left[\left(V_{i1}^2+U_{i1}^2\right)
B_{\chi}^{(d)}(x_{\chi_ig},x_{\tilde{u}_kg},x_{\tilde{u}_hg})
\right.
\nonumber \\
&+&
\left.
2\frac{m_{\chi_i}}{m_g}
V_{i1}U_{i1}B_{\chi}^{(u)}(x_{\chi_ig},x_{\tilde{u}_kg},x_{\tilde{u}_hg})
\right] ,
\label{gbox}
\eea
where $\Delta_{hk}=1$ for $h=k$ and $\Delta_{hk}=-1$ for $h\ne k$.
Finally, the explicit form
of the functions resulting from loop integration is
given in the appendix.

Notice that the supersymmetric box diagrams contribute to the operators
$Q_3$ and $Q_9$ via two different functions $B^{(u)}$ and $B^{(d)}$,
as opposed to the SM case where a single function ($B_{SM}$) appears.
The function $B^{({\tilde{g}})}$ accounts for the contribution of the box
diagrams
with a gluino, a chargino, and two squarks as internal lines. These
diagrams exist also in the limit of flavor-conserving gluino vertices,
in which we are working. In order not to further increase the number
of free parameters, we have chosen to relate the gluino mass
($m_{\tilde{g}}$) to the
$M$ parameter defined in sect. 2 through the GUT relation:
\beq
m_{\tilde{g}}=\frac{\alpha_s(m_{\W})\sin{\theta_{\W}}M}{\alpha(m_{\W})}.
\eeq
This assumption is however inessential,
because the gluino-box contribution to \ep is completely
negligible.

Using the effective Hamiltonian $H_{eff}^{\Delta S=1}$ of \eq{hs1mu},
we recast \eq{epsp} in the form
\beq
\frac{\epsilon^\prime}{\epsilon}=
\mbox{Im}\lambda_t\frac{G_F~\omega}{2|\epsilon|\mbox{Re}A_0}
y_6(\mu)\vev{Q_6(\mu)}_0\Bigl(1-\Omega \Bigr)
\label{epspf2}
\eeq
\beq
\Omega \equiv \Omega_{\eta + \eta^\prime}-
\sum_{\stackrel{i=1,10}{i\ne 6}}
\frac{y_i(\mu)\vev{Q_i(\mu)}_0}{y_6(\mu)\vev{Q_6(\mu)}_0}
+\frac{1}{\omega}\sum_{i=1,10}
\frac{y_i(\mu)\vev{Q_i(\mu)}_2}{y_6(\mu)\vev{Q_6(\mu)}_0},
\eeq
where $\Omega_{\eta +\eta^\prime}$ is the contribution from $\pi -
\eta - \eta^\prime$ mixing, here taken to be
$\Omega_{\eta +\eta^\prime}= 0.25\pm 0.05$.
Following a general convention, we have factored out the dominant
contribution from the operator $Q_6$ and call $\Omega$ the
correction due to the other operators.

In our numerical analysis, we set the renormalization scale $\mu$
at 1 GeV. In order to evaluate \eq{epspf2}, we need the relevant
hadronic matrix elements at that scale:
\beq
\vev{Q_i}_I\equiv \vev{(\pi \pi )_I|Q_i|K}~~~~~~~~I=0,2,
\eeq
where $I$ is the total isospin of the final pion-pion state. In
the future, lattice calculations should provide the most reliable
evaluation of hadronic matrix elements. For the moment we prefer
to follow the approach of ref. \cite{nlob}.

The strategy is to compute the matrix elements $\vev{Q_i}_I$
at the  $m_c$  scale and then, using the QCD evolution, to
obtain the matrix elements at the scale $\mu<m_c$.
The advantages of this method are explained in detail in ref. \cite{nlob}.
Starting at $\mu=m_c$ one finds \cite{nlob}:
\bea
\vev{Q_1}_0&=&-\frac{1}{9}X B_1^{(1/2)},~~~
\vev{Q_2}_0~=~\frac{5}{9}X B_2^{(1/2)},~~~
\vev{Q_3}_0~=~\frac{1}{3}X B_3^{(1/2)},\nonumber \\
\vev{Q_4}_0&=&\vev{Q_3}_0+\vev{Q_2}_0-\vev{Q_1}_0,~~~
\vev{Q_5}_0~=~\frac{1}{3} B_5^{(1/2)}\vev{\overline{Q_6}}_0,\nonumber \\
\vev{Q_6}_0&=&-4\sqrt{\frac{3}{2}}\left[\frac{m_{\KK}^2}{m_s(\mu)+m_d(\mu)}
\right]^2\frac{f_\pi}{k}B_6^{(1/2)},\nonumber \\
\vev{Q_7}_0&=&-\left[\frac{1}{6}\vev{\overline{Q_6}}_0(k+1)-
\frac{X}{2}\right]B_7^{(1/2)},\nonumber \\
\vev{Q_8}_0&=&-\left[\frac{1}{2}\vev{\overline{Q_6}}_0(k+1)-
\frac{X}{6}\right]B_8^{(1/2)},\nonumber \\
\vev{Q_9}_0&=&\frac{3}{2}\vev{Q_1}_0-\frac{1}{2}\vev{Q_3}_0,~~
\vev{Q_{10}}_0~=~\vev{Q_2}_0+\frac{1}{2}\vev{Q_1}_0-\frac{1}{2}\vev{Q_3}_0,
\label{meq0}
\eea
\bea
\vev{Q_1}_2&=&\vev{Q_2}_2=\frac{4\sqrt{2}}{9}XB_1^{(3/2)},~~~
\vev{Q_i}_2~=~0~~~~~~~i=3,...,6,\nonumber \\
\vev{Q_7}_2&=&-\left[\frac{k}{6\sqrt{2}}
\vev{\overline{Q_6}}_0+\frac{X}{\sqrt{2}}\right]B_7^{(3/2)},\nonumber \\
\vev{Q_8}_2&=&-\left[\frac{k}{2\sqrt{2}}
\vev{\overline{Q_6}}_0+\frac{\sqrt{2}}{6}X\right]B_8^{(3/2)},\nonumber \\
\vev{Q_9}_2&=&\vev{Q_{10}}_2=\frac{3}{2}\vev{Q_1}_2 ,
\label{meq2}
\eea
where
\bea
k&=&\frac{\Lambda_{\chi}^2}{m_{\KK}^2-m_{\pi}^2}\simeq 4.55\nonumber \\
X&=&\sqrt{\frac{3}{2}}f_{\pi}\left(m_{\KK}^2-m_{\pi}^2\right)=
3.71\times 10^{-2} \mbox{GeV}^3\nonumber \\
\vev{\overline{Q_6}}_0&=&\frac{\vev{Q_6}_0}{B_6^{(1/2)}}
\eea
and $B_i^{(1/2)}$, $B_i^{(3/2)}$ stand respectively
for the  $\Delta I=1/2,~3/2$ transitions.

These matrix elements are obtained using the vacuum-insertion approximation
and soft-pion theorems, with all of the non-perturbative information
included in the
$B_i$ parameters.
We will use the following values at the scale $\mu=m_c$ \cite{nlob}:
\bea
&&
B_1^{(1/2)}=5.2,~~B_2^{(1/2)}=5.8\pm 1.1,~~B_6^{(1/2)}=1\pm 0.2,
\nonumber \\
&&
B_1^{(3/2)}=0.55,~~B_8^{(3/2)}=1\pm 0.2.
\label{bparam}
\eea
The remaining $B_i$ parameters play only a minor role in the analysis of
\ep and we will set them equal to 1.
%%%%%%%%%%%%%%%%%%%%%%%%%%%%%%%%%%%%%%%%%%%%%%%%%%%%%%%%%%%%%%%%%%%%%%%%%%%
%                         RESULTS
%%%%%%%%%%%%%%%%%%%%%%%%%%%%%%%%%%%%%%%%%%%%%%%%%%%%%%%%%%%%%%%%%%%%%%%%%%%
\section{Results}
It is well known \cite{flra,bur}, \cite{nlob,nlom}
that the SM value of \ep is typically of order
$10^{-4}$ for $m_t=150-190$ GeV, decreases for increasing $m_t$,
and becomes zero for $m_t=200-220$ GeV. This accidental
cancellation occurs because the parameter $\Omega$ in \eq{epspf2}
approaches 1 as the top quark mass reaches these large values.

There are two main sources of uncertainty in the SM prediction for
\epp. One affects the Wilson coefficients and comes from uncertainties
in SM parameters such as the CKM angles, $m_t$, $m_s$, and $\Lambda_{QCD}$.
As measurements on these parameters become more precise, the
calculations of the Wilson coefficients stand on firmer ground.
We summarize in table 2 the SM parameter values (and their relative
errors) chosen for our numerical analysis.

The second source of uncertainty stems from the hadronic matrix
elements, {\it i.e.} from the $B_i$ parameters discussed in sect.~4.
At present lattice calculations seem to be the only tool capable of
significantly reducing errors.

Our goal in this paper is to study the effect of supersymmetry on \epp.
Because of the cancellation mentioned above, the SM value of \ep
is typically smaller than its separate contributions. In this
situation, new-physics effects have the chance to emerge and increase
significantly the total result.

In our analysis we will vary in the experimentally-allowed region
the seven free parameters of the minimal supersymmetric model defined in
sect.~2. The parameters are subject to the
constraints that all charged supersymmetric
particle masses be heavier than $m_Z/2$ and that the common squark mass
satisfy the CDF bound $\tilde{m}>126$ GeV \cite{cdfsusy}.
We also use the tree-level mass bound $M_{H^+}>m_{\W}$ obeyed by the
supersymmetric charged Higgs boson and the bound $\tan \beta >1$
implied by radiative electroweak symmetry breaking.

Since we are interested here in studying the trend of the supersymmetric
corrections, we start by fixing the SM input parameters to the central
values given in table 2. As can be seen from \eq{epspf2}, there are
three sources of new-physics corrections to \epp: the CKM phase
($\mbox{Im}\lambda_t$), the strong penguin ($y_6$), and the
rest of the
electroweak contribution ($\Omega$). Let us first consider the
CKM phase $\delta$, which is determined by the $\epsilon$ parameter,
\eq{epsf}.

By scanning over the supersymmetric parameter space, we obtain the
allowed region of $\delta$ shown  as a function of $m_{\mbox{susy}}$ in
fig.~3a (for $M_{H^+}=80$ GeV) and in fig.~3b (for $M_{H^+}=\infty$),
for $m_t=174$ GeV.
We have defined $m_{\mbox{susy}}$ as the mass of the lightest charged
supersymmetric particle (either the chargino or the stop). With this
definition, it is easy to see from fig.~3 how improvements on the
experimental limits on supersymmetric particle masses ({\it e.g.}
    from LEP 200) will affect the allowed range of $\delta$. The two SM
solutions
for $\cos \delta$ correspond to the minimum values allowed by the
supersymmetric bands in fig.~3b.
In other words, the presence of supersymmetry
has the effect of reducing the value of $\sin \delta$, both for the
solution in the first quadrant ($\cos \delta >0$) and in the second
quadrant ($\cos \delta <0$); the lighter the charged Higgs, the
stronger the reduction. Since \ep is proportional to $\sin \delta$,
this has the effect of decreasing \epp, as discussed below. If the
constraints from $B^0-\bar{B}^0$ mixing and $b\to s \gamma$ are
imposed, only the solution in the first quadrant survives, as shown
in fig.~4. This is true only because we have fixed the SM parameters at
their central values. If all uncertainties are properly taken into account,
some solutions for $\delta$ in the second quadrant can still be found.
In general however a heavy top disfavors solutions with $\cos \delta
<0$, unless $f_B$ is small.

The impact of supersymmetry on the prediction for \ep is illustrated
in fig.~5. This shows the allowed range of \ep as a function of
$m_{\mbox{susy}}$ for $M_{H^+}=80$ GeV (fig.~5a) and $M_{H^+}=\infty$ (fig.~5b)
with $m_t=174$ GeV and with the parameters subject to
the constraints from $B^0-\bar{B}^0$ mixing and $b\to s \gamma$.
In fig.~5a the strong reduction of \ep in the presence
of a relatively light charged Higgs is apparent. This is partly
caused by the decrease in $\sin \delta$ mentioned above. However the
main reason for such small values of \ep is that $\Omega$ can
approach 1, allowing a complete cancellation for values of $m_t$
smaller than that found in the SM. As shown in fig.~5a, \ep can
even reach negative values if $M_{H^+}$ is small enough. The
charged-Higgs effect  on \ep has already been studied by the
authors of ref.\cite{burh} and we confirm here their results.

As discussed above, the chargino contribution has always the effect to
lower $\sin \delta$. However, depending on the choice of the
supersymmetric parameters, the chargino contribution to $\Omega$
can have either sign. As a consequence, \ep in supersymmetry can
also be enhanced with respect to the SM value, as seen in fig.~5.
This enhancement is numerically rather small, at most 40{\%}.
For comparison,
the SM value of \epp, which can be read in fig.~5b in the limit
$m_{\mbox{susy}}\to \infty$, corresponds to the leading-order calculation.
We recall that next-to-leading effects tend to suppress even further
the value of \ep \cite{nlob,nlom}.

In tables 3 and 4 we report the values of the three
different contributions $\sin \delta$, $\Omega$, and $y_6 \vev{Q_6}_0$
corresponding to the minimum and maximum values of \epp.
Notice that the main effect
of supersymmetry resides in $\sin \delta$ and $\Omega$, while
$y_6 \vev{Q_6}_0$ is never reduced by more than 5{\%}. Finally
fig.~6 shows how the prediction for \ep is affected as
the constraints from $B^0-\bar{B}^0$ mixing and $b\to s \gamma$
are removed.

It is interesting to know for which values of the supersymmetric
parameters \ep reaches its minimum and maximum. The charged-Higgs
contribution is maximized for the minimum allowed values of $M_{H^+}$ and
$\tan \beta$, respectively equal to $m_\W$ and 1.
Notice that all charged-Higgs contributions are proportional
to $1/\tan^2\beta$ and therefore rapidly decrease as $\tan \beta$
increases.
The chargino contribution is typically maximized when the chargino and
the lightest stop masses are small ({\it i.e.} of the order of
$m_{\mbox{susy}}$)
and $\tilde{m}\to \infty$. In this limit, flavor symmetry is indeed
maximally broken. Also the chargino contribution minimizes \ep
for $\tan \beta =1$ and maximizes it for $\tan \beta$ in the range 3-5.

Next we want to take into account the experimental and theoretical
errors of the SM input parameters. We have used the following approach.
We first fix the SM parameters to their central values and, for
given $m_{\mbox{susy}}$ and $M_{H^+}$,
we compute the supersymmetric parameters
which respectively minimize and maximize the contribution to \epp.
Then, keeping these values of the supersymmetric parameters fixed, we
generate a large number of configurations using gaussian and flat
distributions for the SM inputs with respectively experimental and
theoretical errors. We obtain
an event distribution for \epp, from which we can compute the
average value and standard deviation. Notice that by following this procedure
we are averaging over solutions with positive and negative $\cos \delta$,
both of which are
allowed when the uncertainties on the SM inputs are taken into account.

The results are shown in tables 5--7, for three values of the top quark mass,
$m_t=150,~174,~190$ GeV, imposing the constraints
     from $B^0-\bar{B}^0$ mixing and $b\to s \gamma$. The effect
of relaxing these constraints is illustrated in table 8.
We observe that the enhancement of \ep in
supersymmetry is indistinguishable from the SM result within
a standard deviation,
even in the most favorable case $m_t=190$ GeV. On the
other hand, the depletion of \ep is substantial and statistically
significant. Unfortunately the experimental sensitivity is presently
not sufficient to identify such small values of \epp.

\section{Conclusions}
We have studied the corrections to \ep in the supersymmetric model with
minimal CP and flavor violation at the leading order in QCD and QED.
Our results are the following.

Supersymmetry can enhance the SM prediction for \ep by at most
40\% for $m_t=174$ GeV and up to 60\% for $m_t=190$ GeV. This
enhancement is of the same order of magnitude as
the SM prediction within one standard deviation
and therefore it is not experimentally distinguishable.
In this respect, minimal supersymmetry cannot explain a value of \ep as
large as the one suggested by the NA31 measurement \cite{na31}. The
enhancement is caused by a reduction of the electroweak factor $\Omega$
and it is attained for chargino and stop masses just beyond the LEP
limit, with all the other squarks and the charged Higgs considerably
heavier.

The most conspicuous effect of minimal supersymmetry is however a strong
depletion of \epp. For certain choices of supersymmetric parameters,
vanishing and even small negative values of \ep can be obtained for the
top quark in the CDF range \cite{cdf}, $m_t=150-190$ GeV. Unfortunately
sensitivity to values of \ep below $10^{-4}$ represents experimentally
a very challenging proposition. This depletion is caused by a reduction of
the value of $\sin \delta$ extracted from $\epsilon$ and by an increase
in the electroweak correction $\Omega$, which can become equal to 1 or
larger. From the short-distance point of view, the dominant corrections
at the origin of the \ep depletion come from a light charged Higgs,
and from charginos and stops just beyond the LEP limit.

\vskip 0.7truecm
{\bf Acknowledgement}\par
One of the authors, E.G. , would like to thank S. Ambrosanio,
M. Ciuchini, and E. Franco for useful discussions and the
Theoretical Physics Division of CERN for its kind hospitality.

%%%%%%%%%%%%%%%%%%%%%%%%%%%%%%%%%%%%%%%%%%%%%%%%%%%%%%%%%%%%%%%%%%%%%%%%%
%                            APPENDIX A
%%%%%%%%%%%%%%%%%%%%%%%%%%%%%%%%%%%%%%%%%%%%%%%%%%%%%%%%%%%%%%%%%%%%%%%%%%%
\section*{Appendix A}
In this appendix we give the expressions for the functions which enter
the penguin and box diagrams.
\par\noindent
%%%%%%%%%%%%%%%%%%%%%%%%%%%%%%%%%%%%%%%%
{\bf Box($\Delta$ S=2 )}
%%%%%%%%%%%%%%%%%%%%%%%%%%%%%%%%%%%%%%%%
\bea
S(x,y)&=&xy\left\{\left[\frac{1}{4}+\frac{3}{2}\frac{1}{1-x}
-\frac{3}{4}\frac{1}{(1-x)^2}\right]\frac{\log x}{x-y}
+\left(x\leftrightarrow y\right)
-\frac{3}{4}\frac{1}{(1-x)(1-y)}\right\}
\nonumber\\
L_1(x,y,z)&=&xy\left[F(x,y,z)+F(y,z,x)+F(z,x,y)\right]
\nonumber\\
L_2(x,y,z)&=&xy\left[xF(x,y,z)+yF(y,z,x)+zF(z,x,y)\right]
\nonumber\\
L_3(x,y,z)&=&\frac{1}{xy}L_2(x,y,z)
\nonumber\\
F(x,y,z)&=&\frac{x\log{x}}{(x-1)(x-y)(x-z)}
\nonumber
\eea
%%%%%%%%%%%%%%%%%%%%%%%%%%%%%%%%%%%%%%%%
\vskip 1truecm
{\bf Box($\Delta$ S=1) }
%%%%%%%%%%%%%%%%%%%%%%%%%%%%%%%%%%%%%%%%
\bea
B_{SM}(x)&=&-\frac{x}{4(x-1)}+\frac{x}{4(x-1)^2}\log{x}
\nonumber\\
B^{u}_{\chi}(x,y,z)&=&-\frac{1}{xy}L_1(x,y,z)
\nonumber\\
B^{d}_{\chi}(x,y,z)&=&L_3(x,y,z)
\nonumber
\eea
%%%%%%%%%%%%%%%%%%%%%%%%%%%%%%%%%%%%%%%%
\vskip 1truecm
{\bf $\gamma$-penguin }
%%%%%%%%%%%%%%%%%%%%%%%%%%%%%%%%%%%%%%%%
\bea
D_{SM}(x)&=&\frac{x^2\left(25-19x\right)}{36\left(x-1\right)^3}+
\frac{\left(-3x^4+30x^3-54x^2+32x-8\right)}{18\left(x-1\right)^4}
\log{x}
\nonumber\\
D_{H}(x)&=&\frac{x\left(47x^2-79x+38\right)}{108\left(x-1\right)^3}+
\frac{x\left(-3x^2+6x-4\right)}{18\left(x-1\right)^4}
\log{x}
\nonumber\\
D_{\chi}(x)&=&\frac{\left(-43x^2+101x-52\right)}{108\left(x-1\right)^3}+
\frac{\left(2x^3-9x+6\right)}{18\left(x-1\right)^4}
\log{x}
\nonumber
\eea
%%%%%%%%%%%%%%%%%%%%%%%%%%%%%%%%%%%%%%%%
\vskip 1truecm
{\bf $Z^0$-penguin }
%%%%%%%%%%%%%%%%%%%%%%%%%%%%%%%%%%%%%%%%
\bea
C_{SM}(x)&=&\frac{x\left(x-6\right)}{8\left(x-1\right)}+
\frac{x\left(3x+2\right)}{8\left(x-1\right)^2}
\log{x}
\nonumber\\
C_{H}(x)&=&-\frac{1}{2}B_{SM}(x)
\nonumber\\
C_{\chi}^{(1)}(x,y)&=&\frac{1}{16\left(y-x\right)}\left[\frac{x^2}{x-1}\log{x}
-\frac{y^2}{y-1}\log{y}\right]
\nonumber\\
C_{\chi}^{(2)}(x,y)&=&\frac{\sqrt{xy}}{8\left(y-x\right)}\left[
\frac{x}{x-1}\log{x}-\frac{y}{y-1}\log{y}\right]
\nonumber
\eea
%%%%%%%%%%%%%%%%%%%%%%%%%%%%%%%%%%%%%%%%
\vskip 1truecm
{\bf Gluon-penguin }
%%%%%%%%%%%%%%%%%%%%%%%%%%%%%%%%%%%%%%%%
\bea
E_{SM}(x)&=&\frac{x\left(x^2+11x-18\right)}{12\left(x-1\right)^3}+
\frac{\left(-9x^2+16x-4\right)}{6\left(x-1\right)^4}
\log{x}
\nonumber\\
E_{H}(x)&=&\frac{x\left(7x^2-29x+16\right)}{36\left(x-1\right)^3}+
\frac{x\left(3x-2\right)}{6\left(x-1\right)^4}
\log{x}
\nonumber\\
E_{\chi}(x)&=&\frac{\left(-11x^2+7x-2\right)}{36\left(x-1\right)^3}+
\frac{x^3}{6\left(x-1\right)^4}
\log{x}
\nonumber
\eea
\newpage
%%%%%%%%%%%%%%%%%%%%%%%%%%%%%%%%%%%%%%%%%%%%%%%%%%%%%%%%%%%%%%%%
%                      BIBLIOGRAPHY
%%%%%%%%%%%%%%%%%%%%%%%%%%%%%%%%%%%%%%%%%%%%%%%%%%%%%%%%%%%%%%%%

\newpage
%%%%%%%%%%%%%%%%%%%%%%%%%%%%%%%%%%%%%%%%%%%%%%%%%%%%%%%%%%%
%%%%%%%%%%%%%%%%%%%% TABLE 1 %%%%%%%%%%%%%%%%%%%%%%%%%%%%%%
% valori delle eta_i nei i box delta s=2
%=======================================
\begin{table}
\centering
\begin{tabular}{|c|c|c|c|c|} \hline
$          $&$ \W\W $&$\HH\HH$&$\HH\W  $&$ \chi $
\\ \hline
$   \eta_1 $&$ 0.85 $&$ 0.07 $&$ 0.21 $&$ 0.62 $
\\ \hline
$   \eta_2 $&$ 0.62 $&$ 0.62 $&$ 0.62  $&$ 0.62 $
\\ \hline
$   \eta_3 $&$ 0.36 $&$ 0.21 $&$ 0.21  $&$ 0.62 $
\\ \hline
%%%%%%%%%%%%%%%%%%%%%%%%%%%% fino qui %%%%%%%%%%%%%%
%%%%%%%%%%%%%%%%%%%%%%%%%%%%%%%%%%%%%%%%%%%%%%%%%%%%%%%%%%%%%%%%%%%%%%%%%
\end{tabular}
\caption[]{ Numerical values for the parameters $\eta_i$ which
represent the QCD corrections to box diagrams.}
\end{table}
%%%%%%%%%%%%%%%%%%%%%%%%%%%%%%%%%%%%%%%%%%%%%%%%%%%%%%%%%%%%%%%
%%%%%%%%%%%%%%%%%%%%%%%%%%%%% TABLE 2 %%%%%%%%%%%%%%%%%%%%%%%%%%%%%%
% valori centrali e dev. stand. dei parametri rilevanti dello SM
%=====================================================================
\begin{table}
\centering
\begin{tabular}{|c|c|c|c|c|c|} \hline
$ B_K $
& $|V_{cb}|$
&$|V_{ub}/V_{cb}|$
&$m_s \left(\mbox{GeV}\right)$
&$\Omega_{\eta+\eta^\prime}$
&$\Lambda^{(4)}_{{\scriptscriptstyle QCD}}\left(\mbox{GeV}\right)$
\\ \hline
$  0.65\pm 0.15 $
&$ 0.040\pm 0.004 $
&$ 0.09\pm 0.04 $
&$ 0.15\pm 0.03 $
&$ 0.25\pm 0.05 $
&$ 0.3\pm 0.1   $
\\ \hline
%%%%%%%%%%%%%%%%%%%%%%%%%%%% fino qui %%%%%%%%%%%%%%
%%%%%%%%%%%%%%%%%%%%%%%%%%%%%%%%%%%%%%%%%%%%%%%%%%%%%%%%%%%%%%%%%%%%%%%%%
\end{tabular}
\caption[]{ Numerical values for the
relevant input parameters of the SM used in $\epsilon$ and
\epp}
\end{table}
%%%%%%%%%%%%%%%%%%%%%%%%%%%%%%%%%%%%%%%%%%%%%%%%%%%%%%%%%%%%%%%
%%%%%%%%%%%%%%%%%%%%%%%%%%%%% TABLE 3 %%%%%%%%%%%%%%%%%%%%%%%%%%%%%%
\begin{table}[t]
\centering
\begin{tabular}{|c|c|c|c|c|c|} \hline
$m_{\mbox{susy}}\left(\mbox{GeV}\right)$
& $M_{H^+}\left(\mbox{GeV}\right)$
& $\sin{\delta}$
& $\Omega$
& $y_6\vev{Q_6}_0\left(\mbox{GeV}^3\right)$
& $\left(\epsilon^\prime /\epsilon\right)_{\mbox{min}}$
\\ \hline\hline
 45      &  80 & 0.567 & $1.08~$  & 2.71$\times10^{-2}$  &
-6.28$\times10^{-5}$
\\ \hline
 45& $\infty$ & 0.567 & 0.784 & 2.64$\times10^{-2}$  &
{}~1.65$\times10^{-4}$
\\ \hline
 60& $\infty$ & 0.553 & 0.721 & 2.69$\times10^{-2}$  &
{}~2.10$\times10^{-4}$
\\ \hline
100& $\infty$ & 0.822 & 0.745 & 2.81$\times10^{-2}$  &
{}~3.00$\times10^{-4}$
\\ \hline
$\infty$      & $\infty$ & 0.978 & 0.682 & 2.85$\times10^{-2}$  &
{}~4.50$\times10^{-4}$
\\ \hline
%%%%%%%%%%%%%%%%%%%%%%%%%%%%%%%%%%%%%%%%%%%%%%%%%%%%%%%%%%%%%%%%%%%%%%%%%
\end{tabular}
\caption[]{ The minimum values of \ep
and the corresponding values of
$\sin{\delta}$, $\Omega$ and $y_6\vev{Q_6}_0$  for different values
of $m_{\mbox{susy}}$ and $M_{H^+}$
for $m_t=174$GeV. The parameters are subject to the constraints
   from $B^0-\bar{B}^0$ mixing and $b\to s \gamma$.}
\end{table}
%%%%%%%%%%%%%%%%%%%%%%%%%%%%%%%%%%%%%%%%%%%%%%%%%%%%%%%%%%%%%%%%%%%%%%%%%
%%%%%%%%%%%%%%%%%%%%%%%%%%%%% TABLE 4 %%%%%%%%%%%%%%%%%%%%%%%%%%%%%%
\begin{table}[t]
\centering
\begin{tabular}{|c|c|c|c|c|c|} \hline
$m_{\mbox{susy}}\left(\mbox{GeV}\right)$
& $M_{H^+}\left(\mbox{GeV}\right)$
& $\sin{\delta}$
& $\Omega$
& $y_6\vev{Q_6}_0\left(\mbox{GeV}^3\right)$
& $\left(\epsilon^\prime /\epsilon\right)_{\mbox{max}}$
\\ \hline\hline
 45      &  80      & 0.798 & 0.446 & 2.78$\times10^{-2}$ & 6.26$\times10^{-4}$
\\ \hline
 45      & $\infty$ & 0.737 & 0.355 & 2.75$\times10^{-2}$  &
6.64$\times10^{-4}$
\\ \hline
 60      & $\infty$ & 0.781 & 0.435 & 2.77$\times10^{-2}$  &
6.24$\times10^{-4}$
\\ \hline
100      & $\infty$ & 0.878 & 0.555 & 2.80$\times10^{-2}$  &
5.57$\times10^{-4}$
\\ \hline
$\infty$ & $\infty$ & 0.978 & 0.682 & 2.85$\times10^{-2}$  &
4.50$\times10^{-4}$
\\ \hline
%%%%%%%%%%%%%%%%%%%%%%%%%%%%%%%%%%%%%%%%%%%%%%%%%%%%%%%%%%%%%%%%%%%%%%%%%
\end{tabular}
\caption[]{ The maximum values of \ep
and the corresponding values of
$\sin{\delta}$, $\Omega$ and $y_6\vev{Q_6}_0$  for different values
of $m_{\mbox{susy}}$ and $M_{H^+}$
for $m_t=174$GeV. The parameters are subject to the constraints
   from $B^0-\bar{B}^0$ mixing and $b\to s \gamma$.}
\end{table}
%%%%%%%%%%%%%%%%%%%%%%%%%%%%%%%%%%%%%%%%%%%%%%%%%%%%%%%%%%%%%%%%%%%%%%%%%
%%%%%%%%%%%%%%%%%%%%%%%%%%%%% TABLE 5 %%%%%%%%%%%%%%%%%%%%%%%%%%%%%%
\begin{table}[t]
\centering
\begin{tabular}{|c|c|c|c|} \hline
 $m_{\mbox{susy}}\left(\mbox{GeV}\right)$
& $M_{H^+}\left(\mbox{GeV}\right)$
& $ \left(\epsilon^\prime /\epsilon \right)_{\mbox{min}}\times10^{4} $
& $ \left(\epsilon^\prime /\epsilon \right)_{\mbox{max}}\times10^{4} $
\\ \hline\hline
 45      &  80      & 1.40$\pm$ 1.30 & 7.24$\pm$ 2.80
\\ \hline
 45      & $\infty$ & 3.32$\pm$ 1.80 & 7.30$\pm$ 2.83
\\ \hline
 60      & $\infty$ & 3.82$\pm$ 2.09 & 7.14$\pm$ 2.80
\\ \hline
100      & $\infty$ & 5.00 $\pm$ 2.43 & 6.83$\pm$ 2.73
\\ \hline
$\infty$ & $\infty$ & 6.49$\pm$ 2.71 & 6.49$\pm$ 2.71
\\ \hline
%%%%%%%%%%%%%%%%%%%%%%%%%%%%%%%%%%%%%%%%%%%%%%%%%%%%%%%%%%%%%%%%%%%%%%%%%
\end{tabular}
\caption[]{ The minimum and maximum average values of
\ep and their standard deviations
for different values
of $m_{\mbox{susy}}$ and $M_{H^+}$
for $m_t=150$GeV. The parameters are subject to the constraints
    from $B^0-\bar{B}^0$ mixing and $b\to s \gamma$.}
\end{table}
%%%%%%%%%%%%%%%%%%%%%%%%%%%%%%%%%%%%%%%%%%%%%%%%%%%%%%%%%%%%%%%%%%%%%%%%%
%%%%%%%%%%%%%%%%%%%%%%%%%%%%% TABLE 6 %%%%%%%%%%%%%%%%%%%%%%%%%%%%%%
\begin{table}[t]
\centering
\begin{tabular}{|c|c|c|c|} \hline
 $m_{\mbox{susy}}\left(\mbox{GeV}\right)$
& $M_{H^+}\left(\mbox{GeV}\right)$
& $ \left(\epsilon^\prime /\epsilon \right)_{\mbox{min}}\times10^{4} $
& $ \left(\epsilon^\prime /\epsilon \right)_{\mbox{max}}\times10^{4} $
\\ \hline\hline
 45      &  80      & -0.432$\pm$ 1.04 & 5.44$\pm$ 2.60
\\ \hline
 45      & $\infty$ & ~1.51$\pm$ 1.14 & 5.68$\pm$ 2.66
\\ \hline
 60      & $\infty$ & ~1.92$\pm$ 1.33 & 5.41$\pm$ 2.57
\\ \hline
100      & $\infty$ & ~2.75$\pm$ 1.81 & 5.03$\pm$ 2.46
\\ \hline
$\infty$ & $\infty$ & ~4.27$\pm$ 2.29 & 4.27$\pm$ 2.29
\\ \hline
%%%%%%%%%%%%%%%%%%%%%%%%%%%%%%%%%%%%%%%%%%%%%%%%%%%%%%%%%%%%%%%%%%%%%%%%%
\end{tabular}
\caption[]{ The minimum and maximum average values of
\ep and their standard deviations
for different values
of $m_{\mbox{susy}}$ and $M_{H^+}$
for $m_t=174$GeV. The parameters are subject to the constraints
     from $B^0-\bar{B}^0$ mixing and $b\to s \gamma$.}
\end{table}
%%%%%%%%%%%%%%%%%%%%%%%%%%%%%%%%%%%%%%%%%%%%%%%%%%%%%%%%%%%%%%%%%%%%%%%%%
%%%%%%%%%%%%%%%%%%%%%%%%%%%%% TABLE 7 %%%%%%%%%%%%%%%%%%%%%%%%%%%%%%
\begin{table}[t]
\centering
\begin{tabular}{|c|c|c|c|} \hline
 $m_{\mbox{susy}}\left(\mbox{GeV}\right)$
& $M_{H^+}\left(\mbox{GeV}\right)$
& $ \left(\epsilon^\prime /\epsilon \right)_{\mbox{min}}\times10^{4} $
& $ \left(\epsilon^\prime /\epsilon \right)_{\mbox{max}}\times10^{4} $
\\ \hline\hline
 45      &  80      & -1.44$\pm$ 1.28 & 4.32$\pm$ 2.23
\\ \hline
 45      & $\infty$ & ~0.61$\pm$ 0.83 & 4.65$\pm$ 2.43
\\ \hline
 60      & $\infty$ & ~0.99 $\pm$ 0.94& 4.39$\pm$ 2.32
\\ \hline
100      & $\infty$ & ~1.48 $\pm$ 1.38 & 3.80 $\pm$ 2.10
\\ \hline
$\infty$ & $\infty$ & ~2.84$\pm$ 1.93& 2.84 $\pm$ 1.93
\\ \hline
%%%%%%%%%%%%%%%%%%%%%%%%%%%%%%%%%%%%%%%%%%%%%%%%%%%%%%%%%%%%%%%%%%%%%%%%%
\end{tabular}
\caption[]{ The minimum and maximum average values of
\ep and their standard deviations
for different values
of $m_{\mbox{susy}}$ and $M_{\HH^+}$
for $m_t=190$GeV. The parameters are subject to the constraints
    from $B^0-\bar{B}^0$ mixing and $b\to s \gamma$.}
\end{table}
%%%%%%%%%%%%%%%%%%%%%%%%%%%%%%%%%%%%%%%%%%%%%%%%%%%%%%%%%%%%%%%%%%%%%%%%%
%%%%%%%%%%%%%%%%%%%%%%%%%%%%% TABLE 8 %%%%%%%%%%%%%%%%%%%%%%%%%%%%%%
\begin{table}[t]
\centering
\begin{tabular}{|c|c|c|c|} \hline
 $m_{\mbox{susy}}\left(\mbox{GeV}\right)$
& $M_{H^+}\left(\mbox{GeV}\right)$
& $ \left(\epsilon^\prime /\epsilon \right)_{\mbox{min}}\times10^{4} $
& $ \left(\epsilon^\prime /\epsilon \right)_{\mbox{max}}\times10^{4} $
\\ \hline\hline
 45      &  80      & -0.440$\pm$ 1.05 & 5.39$\pm$ 2.67
\\ \hline
 45      & $\infty$ & ~1.51$\pm$ 1.18 & 5.62$\pm$ 2.76
\\ \hline
 60      & $\infty$ & ~1.92$\pm$ 1.37 & 5.37$\pm$ 2.67
\\ \hline
100      & $\infty$ & ~2.72$\pm$ 1.81 & 4.90$\pm$ 2.51
\\ \hline
$\infty$ & $\infty$ & ~4.17$\pm$ 2.33 & 4.17$\pm$ 2.33
\\ \hline
%%%%%%%%%%%%%%%%%%%%%%%%%%%%%%%%%%%%%%%%%%%%%%%%%%%%%%%%%%%%%%%%%%%%%%%%%
\end{tabular}
\caption[]{ The minimum and maximum average values of
\ep and their standard deviations
for different values
of $m_{\mbox{susy}}$ and $M_{H^+}$
for $m_t=174$GeV. The parameters are not subject to the constraints
    from $B^0-\bar{B}^0$ mixing and $b\to s \gamma$.}
\end{table}
\newpage
%%%%%%%%%%%%%%%%%%%%%%%%%%%%%%%%%%%%%%%%%%%%%%%%%%%%%%%%%%%%%%%%%%%%%%%%%
%                     FIGURE 1   feynman DIAGRAMS FOR  delta_s=2
%%%%%%%%%%%%%%%%%%%%%%%%%%%%%%%%%%%%%%%%%%%%%%%%%%%%%%%%%%%%%%%%%%%%%%%%%
\begin{figure}
\epsfxsize=16.0cm
\epsffile{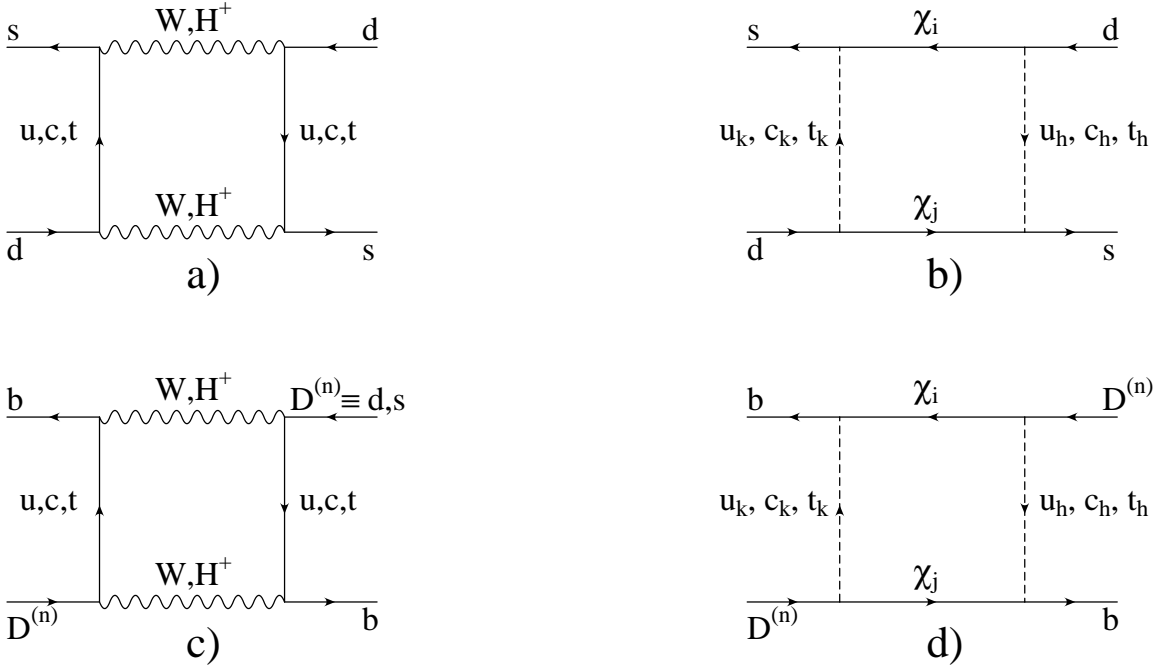}

\caption{The box diagrams $\Delta S=2$ (a-b) and
$\Delta B=2$ (c-d) for $\epsilon$ and $B-\bar{B}$ mixing respectively,
in the minimal supersymmetric SM.}
\end{figure}
\newpage
%%%%%%%%%%%%%%%%%%%%%%%%%%%%%%%%%%%%%%%%%%%%%%%%%%%%%%%%%%%%%%%%%%%%%%%%%
%                     FIGURE 2   feynman  DIAGRAMS FOR  delta_s=1
%%%%%%%%%%%%%%%%%%%%%%%%%%%%%%%%%%%%%%%%%%%%%%%%%%%%%%%%%%%%%%%%%%%%%%%%%
\begin{figure}
\epsfysize=19.0cm
\epsffile{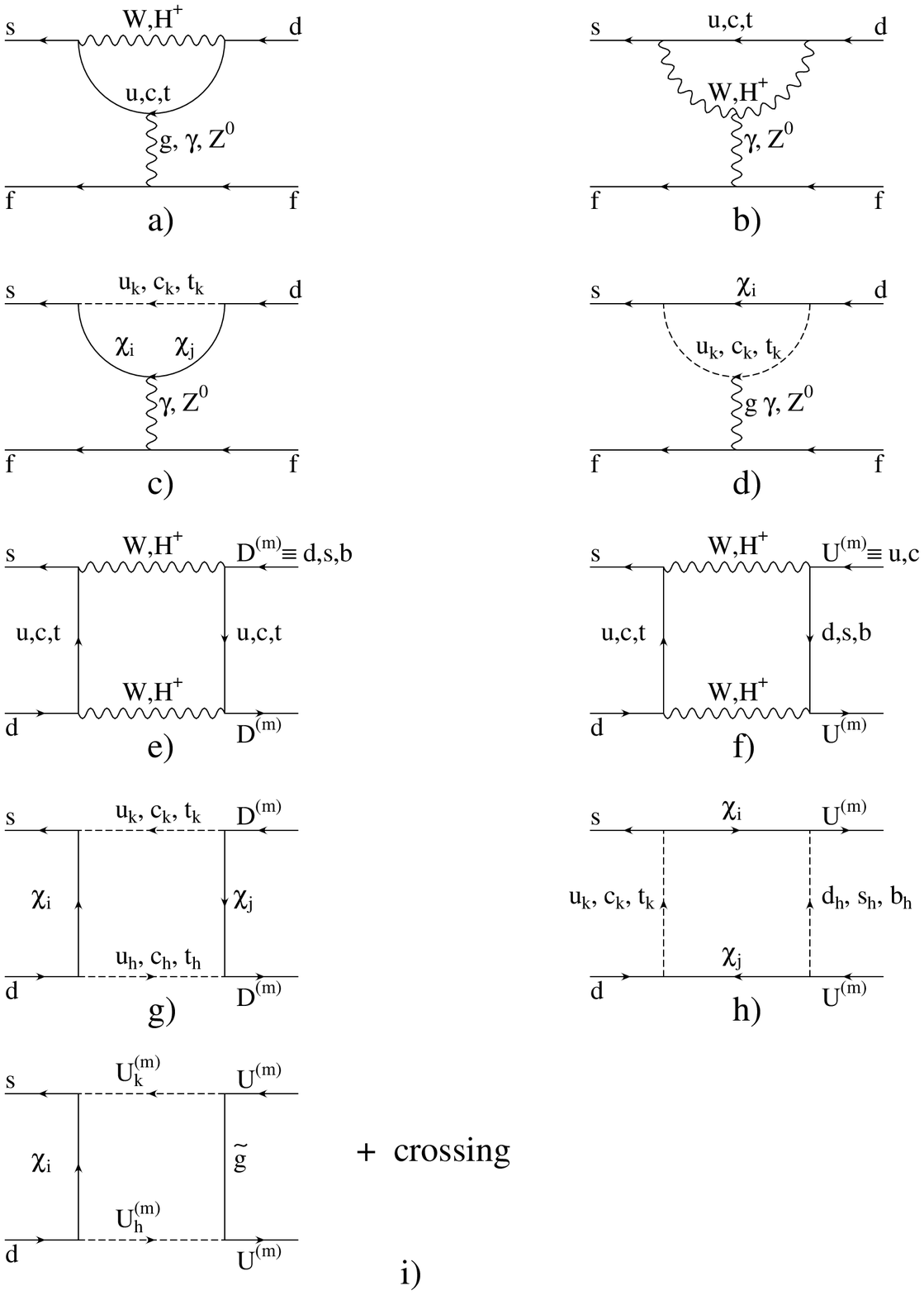}

\caption{The penguin (a-d) and the box (e-i) diagrams for \ep
in the minimal supersymmetric SM.}
\end{figure}
\newpage
%%%%%%%%%%%%%%%%%%%%%%%%%%%%%%%%%%%%%%%%%%%%%%%%%%%%%%%%%%%%%%%%%%%%%%%%%
%                     FIGURE 3A+3B
%%%%%%%%%%%%%%%%%%%%%%%%%%%%%%%%%%%%%%%%%%%%%%%%%%%%%%%%%%%%%%%%%%%%%%%%%
\begin{figure}      % produce figure here
  		    % use [t] for figure at the top of page
		    % use [b] for bottom
		    % use [?] for def choice from LaTeX
    \begin{center}
       \setlength{\unitlength}{1truecm}
       \begin{picture}(12.0,17.0)
          \put(-2.,4.5){\special{fig4_cos.ps}} %mettere nome file .ps qui
          \put(-2.,-4.5){\special{fig3_cos.ps}} %mettere nome file .ps qui
       \end{picture}
    \end{center}
    \caption{ {The bounds of $\cos{\delta}$
as a function of $m_{\mbox{susy}}$ for the charged Higgs mass
$M_{\HH}=80\mbox{GeV}$ and $M_{\HH}=\infty$ respectively in figure a) and b).
The parameters are not subject to the constraints
    from $B^0-\bar{B}^0$ mixing and $b\to s \gamma$.}}
\end{figure}
%%%%%%%%%%%%%%%%%%%%%%%%%%%%%%%%%%%%%%%%%%%%%%%%%%%%%%%%%%%%%%%%%%%%%%%%%
%                     FIGURE 4A+4B
%%%%%%%%%%%%%%%%%%%%%%%%%%%%%%%%%%%%%%%%%%%%%%%%%%%%%%%%%%%%%%%%%%%%%%%%%
\begin{figure}      % produce figure here
  		    % use [t] for figure at the top of page
		    % use [b] for bottom
		    % use [?] for def choice from LaTeX
    \begin{center}
       \setlength{\unitlength}{1truecm}
       \begin{picture}(12.0,17.0)
          \put(-2.,4.5){\special{fig2_cos.ps}} %mettere nome file .ps qui
          \put(-2.,-4.5){\special{fig1_cos.ps}} %mettere nome file .ps qui
       \end{picture}
    \end{center}
    \caption{ {The bounds of $\cos{\delta}$
as a function
of $m_{\mbox{susy}}$ for the charged Higgs mass $M_{\HH}=80\mbox{GeV}$  and
$M_{\HH}=\infty$ respectively in figure a) and b).
The parameters are subject to the constraints
    from $B^0-\bar{B}^0$ mixing and $b\to s \gamma$
and the solution for $\cos{\delta}<0$ is excluded.}}
\end{figure}
\newpage
%%%%%%%%%%%%%%%%%%%%%%%%%%%%%%%%%%%%%%%%%%%%%%%%%%%%%%%%%%%%%%%%%%%%%%%%%
%                     FIGURE 5A+5B
%%%%%%%%%%%%%%%%%%%%%%%%%%%%%%%%%%%%%%%%%%%%%%%%%%%%%%%%%%%%%%%%%%%%%%%%%
\begin{figure}      % produce figure here
  		    % use [t] for figure at the top of page
		    % use [b] for bottom
		    % use [?] for def choice from LaTeX
    \begin{center}
       \setlength{\unitlength}{1truecm}
       \begin{picture}(12.0,17.0)
          \put(-2.,4.5){\special{fig2.ps}} %mettere nome file .ps qui
          \put(-2.,-4.5){\special{fig1.ps}} %mettere nome file .ps qui
       \end{picture}
    \end{center}
    \caption{ {The bounds of \ep with $\cos{\delta}>0$
as a function
of $m_{\mbox{susy}}$ for the charged Higgs mass $M_{\HH}=80\mbox{GeV}$ and
$M_{\HH}=\infty$ respectively in figure a) and b).
The parameters are subject to the constraints
    from $B^0-\bar{B}^0$ mixing and $b\to s \gamma$ and
the solution for $\cos{\delta}<0$ is excluded.}}
\end{figure}
\newpage
%%%%%%%%%%%%%%%%%%%%%%%%%%%%%%%%%%%%%%%%%%%%%%%%%%%%%%%%%%%%%%%%%%%%%%%%%
%                     FIGURE 6A+6B
%%%%%%%%%%%%%%%%%%%%%%%%%%%%%%%%%%%%%%%%%%%%%%%%%%%%%%%%%%%%%%%%%%%%%%%%%
\begin{figure}      % produce figure here
  		    % use [t] for figure at the top of page
		    % use [b] for bottom
		    % use [?] for def choice from LaTeX
    \begin{center}
       \setlength{\unitlength}{1truecm}
       \begin{picture}(12.0,17.0)
          \put(-2.,4.5){\special{fig4.ps}} %mettere nome file .ps qui
          \put(-2.,-4.5){\special{fig3.ps}} %mettere nome file .ps qui
       \end{picture}
    \end{center}
    \caption{ The bounds of \ep
as a function of $m_{\mbox{susy}}$ for the charged Higgs mass
$M_{\HH}=80\mbox{GeV}$ and $M_{\HH}=\infty$ respectively in figure a) and b).
The solid and dashed curves correspond to the solutions
with $\cos{\delta}>0$ and $\cos{\delta}<0$ respectively.
The parameters are not subject to the constraints
    from $B^0-\bar{B}^0$ mixing and $b\to s \gamma$.}
\end{figure}
%___________________________________________________________________________

\begin{thebibliography}{99}
{\small

\bibitem{na31} G.D. Barr \etal ~(NA31 Collaboration), \plb{317}{93}{233}.

\bibitem{e731} L.K. Gibbons \etal ~(E731 Collaboration),\prl{70}{93}{1203}.

\bibitem{nlob} A.J. Buras, M. Jamin, and M.E. Lautenbacher, \npb{408}{93}{209}.

\bibitem{nlom} M. Ciuchini, E. Franco, G. Martinelli, and L. Reina,
\plb{301}{93}{263}.

\bibitem{cdf} F. Abe \etal ~(CDF Collaboration), \prl{73}{94}{225}

\bibitem{burh} G. Buchalla, A.J. Buras, M.K. Harlander, M.E. Lautenbacher,
and C. Salazar, \npb{355}{91}{305}.

\bibitem{epssusy} J.M. Gerard, W. Grimus, and A. Raychaudhuri,
\plb{145}{84}{400};\\
J.M. Gerard, W. Grimus, A. Masiero, D.V. Nanopoulos, and A. Raychaudhuri,
\npb{253}{85}{93};\\
M. Dugan, B. Grinstein, and L. Hall, \npb{255}{85}{413};\\
A. Dannenberg, L. Hall, and L. Randall, \npb{271}{86}{574}.

\bibitem{lo}
J. Bijnens and M.B. Wise, \plb{137}{84}{245};\\
A.J. Buras, and J.M. Gerard, \plb{192}{87}{156};\\
S.R. Sharpe, \plb{194}{87}551;\\
M. Lusignoli, \npb{325}{89}{33}.

\bibitem{nlo} A.J. Buras, M. Jamin, M.E. Lautenbacher, and P.H. Weisz,
\npb{370}{92}{69}, and Addendum, \npb{375}{92}{501};\\
A.J. Buras, M. Jamin, and M.E. Lautenbacher, \npb{400}{93}{37} and\\
\npb{400}{93}{75};\\
M. Ciuchini, E. Franco, G. Martinelli, and L. Reina, \npb{415}{94}{403}.

\bibitem{fcnc} M.J. Duncan, \npb{221}{83}{285};\\
J.F. Donoghue, H.P. Nilles, and D. Wyler, \plb{128}{83}{55}.

\bibitem{cdfsusy} F. Abe \etal ~(CDF Collaboration), \prl{69}{92}{3439}.

\bibitem{kap} L. Ibanez and D. Lust, \npb{382}{92}{};\\
V.S. Kaplunovsky and J. Louis, \plb{306}{93}{269}.

\bibitem{bbmr} S. Bertolini, F.M. Borzumati, A. Masiero, and G. Ridolfi,
\npb{353}{91}{591}.

\bibitem{rpd} Particle Data Group, Review of Particle Properties,
\prd{45}{92}{1}.

\bibitem{arg} H. Albrecht \etal ~(ARGUS Collaboration), \plb{209}{88}{119}.

\bibitem{cleo} M. Artuso \etal ~(CLEO Collaboration), \prl{62}{89}{2233}.

\bibitem{giu} R. Barbieri and G.F. Giudice, \plb{309}{93}{86}.

\bibitem{cleobsg} R. Ammar \etal ~(CLEO Collaboration), \prl{71}{93}{674}.

\bibitem{effec} F. Gilman and M. Wise, \plb{83}{79}{83};\\
B. Guberina and R.D. Peccei, \npb{163}{80}{289};\\
F.J. Gilman and J.S. Hagelin, \plb{126}{83}{111};\\
A.J. Buras and J.M. Gerard, \plb{203}{88}{272}.

\bibitem{flra} J.M Flynn and L. Randall, \plb{224}{89}{221},
Erratum {\bf B 235} (1990) 412;\\
E.A. Paschos and Y.L. Wu, \mpla{6}{91}{93};\\
M. Lusignoli, L. Maiani, G. Martinelli, and L. Reina, \npb{369}{92}{139}.

\bibitem{bur} G. Buchalla, A.J. Buras, and M.K. Harlander, \npb{337}{90}{313}.

\bibitem{bfg} S. Bertolini, M. Fabbrichesi, and
E. Gabrielli, \plb{327}{94}{136};\\
N.G. Deshpande, X.G. He, and S. Pakvasa, \plb{326}{94}{307}.

}
\end{thebibliography}
\end{document}